\documentclass[twocolumn,aps,showpacs,floatfix,amsmath,amssymb,superscriptaddress,prx]{revtex4-2}
\usepackage{amsmath, amsfonts, amssymb, graphics, graphicx, epsfig, color, times, indentfirst, layout, subfigure, multirow, bm, soul}
\usepackage{hyperref}
\usepackage{threeparttable}
\usepackage{graphicx}
\usepackage{dcolumn}
\usepackage{bm}
\usepackage{dsfont}
\usepackage{xcolor}
\usepackage{ulem}
\newcommand{\bc}{\begin{center}}
\newcommand{\ec}{\end{center}}
\newcommand{\be}{\begin{equation}}
\newcommand{\ee}{\end{equation}}
\newcommand{\ba}{\begin{array}}
\newcommand{\ea}{\end{array}}
\newcommand{\beq}{\begin{eqnarray}}
\newcommand{\eeq}{\end{eqnarray}}
\newcommand{\ket}[1]{\left| {#1}\right\rangle}

\newcommand{\e}[1]{\langle {#1}\rangle}
\newcommand{\abs}[1]{\left| {#1}\right|}
\newcommand{\trip}[3]{\langle {#1}|{#2}|{#3}\rangle}

\begin{document}

\title{Probing entanglement dynamics and topological transitions on noisy intermediate-scale quantum computers}

\author{Huai-Chun Chang}
\affiliation{Graduate Institute of Applied Physics, National Chengchi University, Taipei 11605, Taiwan}
\affiliation{Physics Division, National Center for Theoretical Sciences, Taipei 10617, Taiwan}
\author{Hsiu-Chuan Hsu}\email{hcjhsu@nccu.edu.tw}
\affiliation{Graduate Institute of Applied Physics, National Chengchi University, Taipei 11605, Taiwan}
\affiliation{Department of Computer Science, National Chengchi University, Taipei 11605, Taiwan}
\author{Yu-Cheng Lin}\email{yc.lin@nccu.edu.tw}
\affiliation{Graduate Institute of Applied Physics, National Chengchi University, Taipei 11605, Taiwan}

\date{\today}
\begin{abstract}
        We simulate quench dynamics of the Su-Schrieffer-Heeger (SSH) chain on the IBM quantum computers,
calculating the R\'enyi entanglement entropy, the twist order parameter and the Berry phase. 
The latter two quantities can be deduced from a slow-twist operator defined in the Lieb-Schultz-Mattis theorem.
The R\'enyi entropy is obtained using a recently developed randomized measurement scheme.
The twist order parameter and the Berry phase 
are measured without the need for additional gates or ancilla qubits.
We consider quench protocols in which a trivial initial state evolves dynamically in time under
the topological SSH Hamiltonian in the fully dimerized limit (the flat-band limit).
During these quenches, there are persistent and periodic oscillations in the time evolution of both entanglement entropy
and twist order parameter.
Through the implementation of error mitigation techniques using a global depolarizing ansatz and postselection, 
our simulations on the IBM devices yield results that closely match exact solutions.
\end{abstract}
\maketitle

\section{Introduction}
\label{intro}
Nonequilibrium dynamics of quantum many-body systems
is an active field of research both in condensed matter physics and quantum information science.
Most of the efforts have focused on dynamics generated by a quantum quench, i.e. 
a sudden change of parameters in the Hamiltonian,  
because it poses important 
fundamental questions that can be investigated in controlled experiments~\cite{Mitra}. 
Among experimental systems, ultracold atoms have proven to be the ideal arena 
to realize systems' Hamiltonians and simulate quantum dynamics using
the inherent quantum mechanical properties~\cite{ColdAtom1,ColdAtom2}.
Recently, cold-atom systems have been further developed to programmable devices~\cite{Analog}
that embody Richard Feynman's proposal for an analog quantum computer~\cite{Feynman}.

Here we focus on quantum quench dynamics implemented by another type of quantum
computing approach, namely digital quantum computing, or called gate-based
quantum computing.  In the digital approach, quantum simulations are carried
out by sequences of quantum gates applied to a small number of physical qubits,
mimicking classical computations.  One key advantage of digital quantum
computing is the versatility and universality it offers, as any operation can
be expressed as a finite set of gates~\cite{Nielson2011}.   However, gate
errors and noise often limit the depth of a quantum circuit where layers of the
required quantum gates are built.  Moreover, one may need to use the Trotter
decomposition~\cite{Trotter} to approximately decompose the unitary time
evolution into a product of one or two-qubit gates while simulating the quantum
dynamics of a many-body system in the digital approach, which generates
additional systematic errors.  To obtain the best possible results from today’s
noisy intermediate-scale quantum (NISQ) devices, numerous techniques for error
correction~\cite{Correction_PRL2022,Correction_NC2023}, error
mitigation~\cite{Cai_rev} and optimization
strategies~\cite{Compression_PRA2022} have been developed in the past few
years.

Several studies have demonstrated the potential of digital quantum computing
for quantum many-body simulations~\cite{qsim_rev}.  In particular, various
entanglement properties, such as entanglement spectra~\cite{Choo},
symmetry-resolved entropies~\cite{Azses}, the second-order R\'enyi
entropy~\cite{Mitigation_PRE,LGT_PRXQ}, for many-body ground states and
dynamical states have been actively explored on current quantum computers.
There is also an increasing interest in probing topological properties with
NISQ devices; recent works include measuring string order parameters crossing
symmetry protected topological phase transitions~\cite{String}, simulating
chiral topological dynamics~\cite{Chiral_dyn}, and probing fractional quantum
Hall states~\cite{FQH}.

In this work, we simulate quench dynamics of the Su-Schrieffer-Heeger (SSH)
chain~\cite{SSH} on IBM-Q devices. The SSH model serves as a fundamental system
for describing a topological insulator that exhibits transitions between
topologically distinct phases in its quench
dynamics~\cite{Chang,Ueda,Yang,Hsu}. Previous simulations of the SSH model and
related models on NISQ devices have predominantly focused on Bloch states in
momentum space~\cite{Guo,Murta,Kemper} or edge states~\cite{Mei,Lee_NPJ}.  Here
we focus on topological properties of the entire system in real space, with up
to 12 qubits.  We quantify the entanglement using the second-order R\'enyi
entanglement entropy and probe topological properties using the twist operator
appearing in the Lieb-Schultz-Mattis theorem~\cite{Lieb,Twist}.  We consider a
special case where the dynamics of entanglement and topology for an arbitrary
system size exhibits persistent and periodic oscillations.  We obtain
analytical closed-form solutions for the dynamical entanglement entropy
starting from two different initial states.  The simulations can be executed
without Trotter errors and without using variational approaches, enabling us to
evaluate the ability of NISQ devices to simulate quantum many-body dynamics
through comparisons with exact solutions. The persistent and periodic
oscillations in entanglement entropy, along with their analytical expressions,
could potentially be used as a tool for calibrating quantum hardware and
benchmarking quantum algorithms designed to compute this quantity.

The paper is organized as follows:
In Sec.~\ref{sec:model}, we define the model and the quench protocols considered in our simulations;
we summarize some known properties of the quench dynamics. In Sec.~\ref{sec:simulation}, we describe
our simulation setups on the IBM-Q devices and present our simulation results, focusing on the R\'enyi entanglement entropy 
and twist order parameters.
We conclude in Sec.~\ref{sec:summary} with a summary and discussion.    
A derivation of the entanglement entropy is presented in Appendix~\ref{append}.

\section{The Model and quench protocols}
\label{sec:model}
We consider the spin-1/2 XX chain of $L=2N$ spins with alternating bond strength, defined by the Hamiltonian:
\begin{eqnarray}
        H_{\text{XX}}&=&\sum_{j=1}^{N} J \left(
       \sigma^{x}_{2j-1}\sigma^{x}_{2j} +\sigma^{y}_{2j-1}\sigma^{y}_{2j}
        \right)\nonumber\\
        &+& \sum_{j=1}^{N-1} J' \left(
        \sigma^{x}_{2j}\sigma^{x}_{2j+1} +\sigma^{y}_{2j}\sigma^{y}_{2j+1}
        \right)\\
        &+& J'_{2N} \left(
        \sigma^{x}_{2N}\sigma^{x}_{1} +\sigma^{y}_{2N}\sigma^{y}_{1}
        \right)\nonumber
        \label{eq:xx}
\end{eqnarray}
where $\sigma_j^{\alpha}$ ($\alpha=x,y$) is the $\alpha$-component of the  Pauli matrix at site $j$;
$J$ and $J'$ are coupling constants on odd and even links, respectively.
The coupling constant on the link $j=2N$ is $J'_{2N}=J'$ for periodic boundary conditions (PBC),
and is zero ( $J'_{2N}=0$) for open boundary conditions (OBC).
The system consists of two sublattices $A$ and $B$, where
sublattice $A$ contains all even sites and sublattice $B$ contains all odd sites.

%%%% L=2N %%%%%%%%%%%%

The XX spin chain is effectively a non-interacting fermion model~\cite{Lieb}.
Using the Jordan–Wigner transformation, 
the dimerized XX model defined above
can be mapped to the SSH model,
which is a half-filled chain of free fermions with alternating hopping described as
\begin{eqnarray}
   H_{\text{SSH}} &=& \sum_{m=1}^{N-1} \left(2J a_m^\dag b_m + 2J' b_m^\dag a_{m+1} + \text{h.c.}\right)\nonumber\\
                  &+& 2J a_N^\dag b_N - (-1)^{\mathcal{N}} 2J'  b_N^\dag a_{1} + \text{h.c.},
\end{eqnarray}
where $m$ indexes $N$ unit cells that are pairs of two sites, one on sublattice $A$ and one on $B$.
The operator $a^\dag_m$($b^\dag_m$) creates a spinless fermion at the $A$($B$)-sublattice site in the $m$th unit cell,
and $\mathcal{N}$ is the the number of fermions.
The constants $2J$ and $2J'$ are regarded as the intracell and intercell hopping amplitudes, respectively. 
The boundary term is affected by the parity of the number of fermions~\cite{Lieb};
depending on evenness or oddness of $\mathcal{N}$, the boundary conditions of the fermionic chain
are antiperiodic or periodic.

We consider the case with PBC and use the Fourier transform of the form:
\be
   \begin{pmatrix}
       a_m\\
       b_m
   \end{pmatrix} = \frac{1}{\sqrt{N}} \sum_k e^{ikm} 
   \begin{pmatrix}
       a_k\\
       b_k
   \end{pmatrix}\,.
\ee
The Hamiltonian in momentum space can be written as
\be
   H_{\text{SSH}} = \sum_k \Psi^\dag_k H_k \Psi_k\,,
   \label{eq:Hk}
\ee
with $\Psi^\dag_k=(a^\dag_k,  b^\dag_k)$ and a single-particle Bloch  Hamiltonian $H_k=\mathbf{d}_k\cdot\bm{\sigma}$,
where
$\bm{\sigma}$ is the Pauli matrix and the vector $\mathbf{d}_k=(d_k^x,d_k^y,d_k^z)$ is given by
\be
             \mathbf{d}_k=(2J+2J'\cos(k),\,2J'\sin(k),\,0)\,.
    \label{eq:d}
\ee
The eigenvalues of the two-level Bloch Hamiltonian are given by 
\be
       E_{\pm}=\pm\abs{\mathbf{d}_k} =\pm2\sqrt{J^2+2JJ'\cos(k)+J'^2}\,,
       \label{eq:Ek}
\ee
and the corresponding eigenvectors are
\be
       \ket{\psi_\pm} = \frac{1}{\sqrt{2}} 
                        \begin{pmatrix}
                           \pm e^{-i\phi(k)} \\
                           1 
                        \end{pmatrix} 
\ee 
with $\phi(k)=\tan^{-1}(d_k^y/d_k^x)$.
The dispersion relation in Eq.(~\ref{eq:Ek}) describes a band structure with a gap $\Delta_E=4\abs{J-J'} $ at $k=\pm \pi$,
as long as $J\neq J'$. In the fully dimerized cases ($J>0$, $J'=0$, or vice versa $J=0$, $J'>0$), the  band structure
becomes flat, i.e. independent of the wavenumber $k$. In this flat-band limit, there is also no difference between 
the periodic and antiperiodic boundary conditions.
The  topological  properties of the system
are encoded in the closed curve that the vector $\mathbf{d}_k$ traces out as $k$ changes from $0$ to $2\pi$.
For $J<J'$, the curve encloses the origin $\mathbf{d}_k=\mathbf{0}$ and the system is in the topological phase;
for $J>J'$, the origin is completely outside the closed curve, corresponding to a trivial phase.
At the critical point $J=J'$ where the band gap closes, the curve  passes through the origin. 
The number of times  that the closed curve goes around the origin
is referred to as the winding number $\nu$,
defining a topological invariant. Other useful tools to characterize topological properties include entanglement measures~\cite{Sirker,Chang,Ueda}
and the Berry phase~\cite{Berry,Zak,Bloch,AIII}.

A transition between two distinct topological phases can occur without
crossing a critical point, but rather by breaking chiral symmetry. This
phenomenon arises when the vector $\mathbf{d}_k$ is no longer confined to the ($d_k^x, d_k^y$)-plane
and instead possesses a finite $d_k^z$ component. Such a scenario is observed in
nonequilibrium dynamics following a sudden quench of certain Hamiltonian
parameters~\cite{Chang,Ueda}, even when $d_k^z$  is absent in both pre-quench and
post-quench Hamiltonians. Here we focus on quenches with the flat-band Hamiltonian with $(J,J')=(0,1)$.
%and PBC.
Two types of initial (pre-quench) states are considered: (i) a classical N\'eel state  described by
\be
       \ket{\psi_\text{N}}=\bigotimes_{j=1}^N \ket{1_{2j-1} 0_{2j}}\,,
       \label{eq:psi_N}
\ee
in terms of the standard basis of $z$ spin components: $\sigma^z\ket{0}=\ket{0}$,
$\sigma^z\ket{1}=-\ket{1}$,
and (ii) a fully dimerized state described by a product of intracell singlets written as
\be
      \ket{\psi_\text{S}}=\bigotimes_{j=1}^N \frac{1}{\sqrt{2}}\left(\ket{0_{2j-1} 1_{2j}} - \ket{1_{2j-1}0_{2j}} \right)\,,
      \label{eq:psi_S} 
\ee
which is equivalent to the ground state of the trivial flat-band  Hamiltonian with $(J,J')=(1,0)$.
For a quench from $\ket{\psi_\text{N}}$, the post-quench state will reach a winding number $\nu=1$ at
times $t_\text{N}=(n-\frac{1}{2})\frac{\pi}{4}$ with $n\in\mathbb{Z}^+$; for a quench from $\ket{\psi_\text{S}}$, the post-quench state will have a winding number $\nu=2$
at $t_\text{S}=(n-\frac{1}{2})\frac{\pi}{2}$~\cite{Chang,Ueda}.
Below we denote by $t_\text{N}^*$ and $t_\text{S}^*$ those time points at which a finite winding number occurs.  

In real space, 
the  
periodic generation of the winding number  manifests itself in the dynamics of 
intercell entanglement formation~\cite{DisorderAIII}. For the initial $\ket{\psi_\text{N}}$ state, the post-quench state 
consists of a set of nearest neighbor intercell entangled states at $t=t_\text{N}^*$; For the initial $\ket{\psi_\text{S}}$ state, 
the post-quench state develops entangled states connecting $A$ and $B$ sublattices in two next-nearest-neighbor cells
(i.e. extended to three lattice spacings) at times $t=t_\text{S}^*$.
The  development of such local entanglement is evident from the time-dependent Wannier functions in
the free-fermion system, as discussed in Appendix~\ref{append}.

\subsection{Entanglement entropy}

\begin{figure}[tb!]
\centerline{\includegraphics[width=8.6cm]{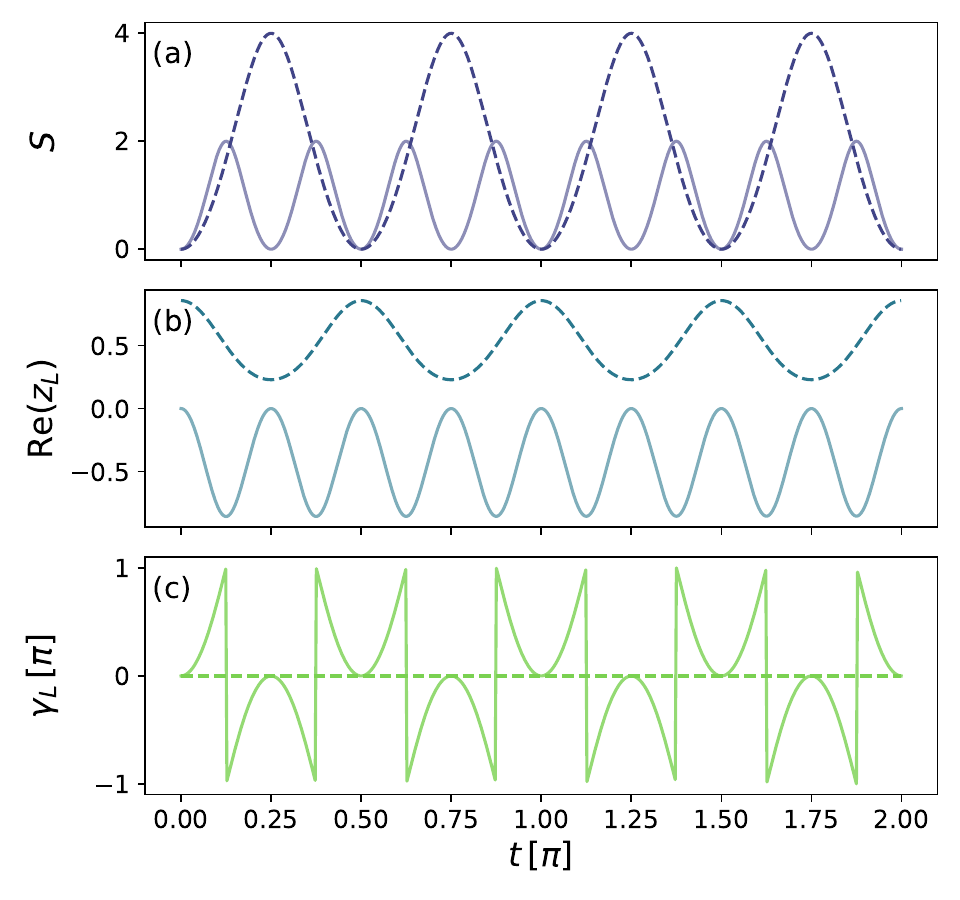}}
%\vskip-2mm
\caption{Exact data for quenches from a N\'eel state $\ket{\psi_\text{N}}$ (solid lines) and from a fully dimerized state $\ket{\psi_\text{S}}$ (dashed lines), driven by the flat-band Hamiltonian with $(J=0, J'=1)$ and periodic boundary conditions. 
The panels from top to bottom show the evolution of 
(a) the R\'enyi entanglement entropy with a symmetric bipartition (given by Eqs.~(\ref{eq:S01}) and (\ref{eq:S02})), 
(b) the twist order parameter for $L=16$, and 
(c) the Berry phase for $L=16$, obtained by exact diagonalization.
}
\label{fig:l16}
%\vskip-3mm
\end{figure}

The formation of intercell entanglement in the post-quench states and their dynamics can be probed by entanglement entropy (EE).
Each maximally entangled state that connects a block of contiguous sites with the rest of the system 
contributes one unit to the block EE (both R\'enyi and von Neumann entropies).
In our settings, the post-quench state at times $t^*_\text{N(S)}$ evolves to a product of maximally entangled states,
the EE of a segment embedded in the chain is then just the number of entangled states that connect sites inside to 
sites outside the segment.
For a closed chain divided into two subsystems, the maximum EE for the state $\ket{\psi_\text{N}(t)}$ 
at $t^*_\text{N}$ is 2, while it is 4 for the state $\ket{\psi_\text{S}(t)}$ at $t^*_\text{S}$.
To observe the time evolution of entanglement, 
we consider the second-order R\'enyi EE, defined as
\be
        S=-\log_2 \text{Tr}[\rho_I^2]\,,
       \label{eq:S}
\ee
where $\rho_I$ is the reduced density matrix %for the first subsystem. 
for a subsystem (denoted by $I$) in the bipartite system.
Using the correlation matrix approach proposed in Ref.~\cite{Peschel},
we can obtain the R\'enyi EE via
\be
      S=-\sum_i \log_2 \left[1-2\xi_i(1-\xi_i) \right]\,,
      \label{eq:S_cor}
\ee
where $\xi_i$ is an eigenvalue of the correlation matrix.
We consider a closed chain of length $L=4\ell$ with an integer $\ell\ge 1$ and a symmetric bipartition such that 
each subsystem contains $\ell$ complete unit cells.
For a quench from $\ket{\psi_\text{N}}$,
the  dynamical R\'enyi EE of a half chain  is
\be
     S^{0\to 1}(t)=-2\log_2\left[1-\sin^2(2\abs{\mathbf{d}_k}t)/2 \right]\,.
     \label{eq:S01}
\ee
where the superscript $0\to 1$ indicates the change in the winding number under the quench. 
For a quench from the fully dimerized state $\ket{\psi_\text{S}}$ in which a transition to $\nu=2$ occurs,
the half-chain R\'enyi EE for a chain with $\ell>1$ evolves as
\be
     S^{0\to 2}(t)=-4\log_2\left[1-\sin^2(\abs{\mathbf{d}_k}t)/2 \right]\,.
     \label{eq:S02}
\ee
We note that $S^{0\to 2}(t)=2S^{0\to 1}(t/2)$ [see Fig.~\ref{fig:l16}(a)].
For a short closed chain of length $L=4$ with only two unit cells, the dynamical R\'enyi EE is reduced to $S^{0\to 1}(t)$.
Technical details of the derivation of the dynamical R\'enyi EE are given in Appendix~\ref{append}.

We note that quasiparticle excitations, which typically contribute to a linear growth of the EE 
after a quench~\cite{Calabrese}, are confined with a group velocity of zero in our flat-band model.
Thus, the persistent oscillations in the EE indicate the absence of relaxation following quenches~\cite{Confinement_nphys}.

%%%%%%%%%%%%%%%%% FIG1 %%%%%%%%%%%%%%%%%%%%%%%
%%%%%%%%%%%%

\subsection{Twist operator}
A unitary operator, called the twist operator, defined via the spin operator $S^z$ ($S^z=\sigma^z/2$ for spin-1/2) 
as
\be
             U_z =  \exp\biggl(i\frac{2\pi}{L} \sum_{j=1}^L j S_j^z\biggr)\,,
             \label{eq:twist}
\ee
for a spin chain of length $L$,
is a useful tool to characterize various topology properties.
This unitary operator
was first introduced in the Lieb-Schultz-Mattis (LSM) theorem~\cite{Lieb,Affleck}, which states that
the amplitude, dubbed as the twist order parameter, given by
\be
            z^{(q)}_L= \trip{\psi}{U^q_z}{\psi}
            \label{eq:zL}
\ee
in terms of the ground state of the XXZ spin chain
can be seen as an indicator of a gapped or gapless excitation spectrum,
depending on whether $z^{(q)}_L$ is nonvanishing or zero in the thermodynamic limit $L\to\infty$.
A similar operator was introduced to spinless electron systems~\cite{Position,Oshikawa_Z}:
\be
          U_n = \exp\biggl(i\frac{2\pi}{L} \sum_{j=1}^L jn_j \biggr)\,,
          \label{Un}
\ee  
where $n_j=S_j^z+1/2$ is the particle number operator at site $j$.
Similar to the twist order parameter $z_L$ defined in Eq.~(\ref{eq:zL}), the ground-state expectation value
\be
        z^{(q)}_N = \trip{\psi}{U^q_n}{\psi}\,,
        \label{eq:zn}
\ee
is an order parameter that
distinguishes an insulating phase and a gapless conducting phase at rational filling $n=p/q$
(where $p/q$ is an irreducible fraction)~\cite{Aligia_PRL},
depending on
whether $z^{(q)}_N\neq 0$ or $z^{(q)}_N = 0$ in the limit of $L\to\infty$~\cite{Localization}.
Both definitions of the twist operator are consistent with the periodic boundary conditions of the chain.

The application of the twist order parameters  have been extended to the study
of various gapped states. 
For instance, the amplitude $z_L$
was proposed as an order parameter to distinguish different types of 
valence-bond-solid(VBS)-like dimerization~\cite{Twist},
even in random and inhomogeneous systems~\cite{Spin2}.
In the case of a spin-1/2 VBS state,
a dimerized pattern of two-spin singlets (valence bonds) formed on odd
links (i.e. intracell links) is a dimer phase called
$(1,0)$-phase, with $z_L^{(1)}=1$ for $L\to\infty$.
The fully dimerized initial state, $\ket{\psi_\mathrm{S}}$, we consider here 
corresponds to a $(1,0)$-phase.
On the other hand, a dimer phase where two-spin singlets are formed on
even links (i.e. intercell links), 
referred to as the $(0,1)$-phase, results in $z_L^{(1)}= -1$ for $L\to\infty$.
This occurs because a $2\pi$ rotation induces a negative sign
for a valence bond at the boundary $(j=L)$.

Here we apply the twist operator $U_z$ to the dynamical states in our quench
protocols. In Fig.~\ref{fig:l16}(b), we demonstrate results for time-dependent
$z_L^{(1)}$ with $L=16$, obtained by exact diagonalization.  We observe a
periodic recurrence of the $(1,0)$-phase in the time-dependent $z_L^{(1)}$
during a quench from the fully dimerized state $\ket{\psi_\mathrm{S}}$.
This dimerization is characterized by the peaks of $z_L^{(1)}$,
with a peak value $[\cos(\pi/L)]^{L/2}$ for a finite chain of length $L$ with PBC~\cite{Twist}.
Interestingly, the twist order parameter can also be applied to the quench from
the N\'eel state $\ket{\psi_\mathrm{N}}$, where a dimerized phase emerges
periodically at $t=t^*_\text{N}$. This dimerized phase is similar to a
$(0,1)$-phase and can be identified by the negative peaks of $\text{Re}(z_L)$, corresponding to $\text{Re}(z_L)=-[\cos(\pi/L)]^{L/2}$.  
It is worth noting that, for the dynamical state
$\ket{\psi_\text{N}(t)}$, the twist order parameter is not purely real, in
contrast to $\ket{\psi_\text{S}(t)}$. This difference arises due to the
breaking of inversion symmetry in $\ket{\psi_\mathrm{N}(t)}$.

Another application of the twist operator that we consider here is its
connection to the Berry phase. In the free fermion model, 
the Berry phase represents the phase angle associated with the evolution of the quantum state 
as each particle undergoes a momentum shift of $2\pi/L$ induced by the twist operator.
Moreover, the definition of the twist operator in Eq.~(\ref{Un}) provides a means to calculate
the expectation value of particle position, $X=\sum_j j n_j$, in an $N$-particle system of size $L$ with PBC at rational filling $N/L=p/q$
through~\cite{Position,Localization,Aligia_PRL}
\be
     \e{X}=\frac{L}{2\pi q} \text{Im}\ln[z_N^{(q)}]\,.
\ee
%at rational filling $N/L=p/q$. 
Therefore, the Berry
phase is directly related to 
the position expectation value~\cite{Position,Localization,Aligia_PRL}:
%the Wannier center,
\be
     %\gamma_L=\frac{L\e{X}}{2\pi q}\,,
      \gamma_L=\text{Im}\ln[z^{(q)}_N]=\frac{2\pi q}{L}\e{X}\,.
      \label{eq:berry}
\ee
%which characterizes the spatial distribution of electronic charge.
This, in turn, establishes a link between the Berry phase and the macroscopic
polarization of a many-electron system: $P=-e\gamma_L/2\pi$, where $e$ is the electron charge~\cite{Position,Localization}.

We apply $U^q_n$ with $q=2$ to our dynamical states at half filling ($n=1/2$).
As shown in Fig.~\ref{fig:l16}(c) The Berry phase for $\ket{\psi_\text{N}(t)}$ jumps between $-\pi$ and $\pi$ at
$t=t_\text{N}^*$, indicating a quantized dynamical Chern number $C=\pm 1$; this result agrees with
previous studies based on the single-particle Hamiltonian in momentum space~\cite{Hsu}.
On the other hand, the Berry phase for $\ket{\psi_\text{S}(t)}$ remains zero for all $t$, reflecting that
the dynamical state has inversion symmetry and the twist order parameter remains real.

\section{Simulations}
\label{sec:simulation}
In this section we present the relevant setups for our simulations on
the IBM quantum computers, and show simulation results for dynamical entanglement entropy
and the time-dependent twist order parameter. 
\subsection{Time evolution}

%%%%%%%%%%%%%%%%% FIG2 %%%%%%%%%%%%%%%%%%%%%%%
\begin{figure}[tb!]
\centerline{\includegraphics[width=8.6cm]{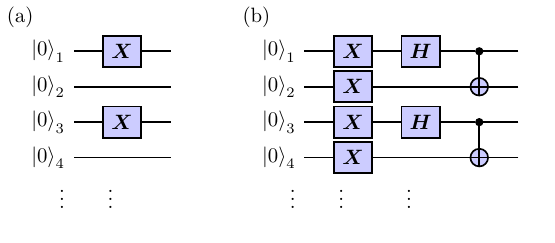}}
%\vskip-2mm
\caption{Quantum gates for preparing the initial states: (a) the N\'eel state $\ket{\psi_\text{N}}$ and
(b) the fully dimerized state with intracell singlets $\ket{\psi_\text{S}}$.}
\label{fig:ini}
%\vskip-3mm
\end{figure}
%%%%%%%%%%%%

The default state on the IBM devices is a tensor product of $\ket{0}$
and the computational basis is also composed by the tensor product of the eigenstates of $\sigma^z$. 
We set up the initial states $\ket{\psi_\text{N}}$ and $\ket{\psi_\text{S}}$ for
our quench protocol using the Pauli gate $\bm{X}$, the Hadamard gate $\bm{H}$ and the controlled-not (CNOT) gate $\bm{C_X}$,
as shown in Fig.~\ref{fig:ini}, where a two-spin singlet state is realized by a set of quantum gates 
$(\bm{C_X})(\bm{H}\otimes\bm{I})(\bm{X}\otimes\bm{X})$.
%$(\bm{X}\otimes\bm{X})(\bm{H}\otimes\bm{I})(\bm{C_X})$.

%%%%%%%%%%%%%%%%% FIG3 %%%%%%%%%%%%%%%%%%%%%%%
\begin{figure}[tb!]
\centerline{\includegraphics[width=8.6cm]{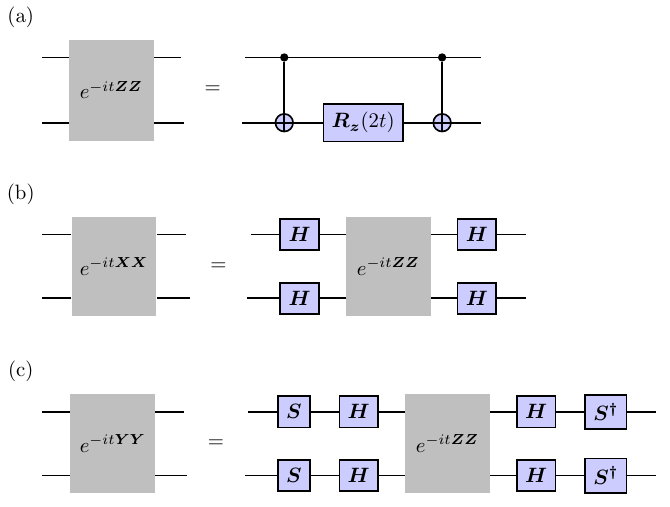}}
%\vskip-2mm
\caption{Quantum circuits to implement the time-evolution operators $U^{zz}(t), U^{xx}(t)$ and $U^{yy}(t)$.
}
\label{fig:exp}
%\vskip-3mm
\end{figure}
%%%%%%%%%%%%

For the post-quench Hamiltonian, the fully dimerized XX chain with $(J,J')=(0,1)$,  
that we consider here, the quantum circuit for the time evolution operator $U(t)=\exp(-iHt)$ with an arbitrary time duration $t$ 
can be constructed without any Trotter error.
The time evolution operator can be split into 
\be
             %U(t) = \prod_j U_{2j}(t) 
             U(t)=\bigotimes_j U_{2j}(t) 
\ee
with
\be
    \begin{split}
              U_{2j}(t) &=\exp\left[-it (\sigma_{2j}^x \sigma_{2j+1}^x + \sigma_{2j}^y \sigma_{2j+1}^y)  \right] \\
                              &=\exp(-it \sigma_{2j}^x \sigma_{2j+1}^x) \exp(-it \sigma_{2j}^y \sigma_{2j+1}^y)\,,
    \end{split}
\ee
where the commutator identity $[\sigma_{j}^x \sigma_{j'}^x,\, \sigma_{j}^y \sigma_{j'}^y]=0$ is applied.
The quantum circuits required for  the time evolution are then the gates to implement the operators $U^{xx}(t)=e^{-it \bm{X} \bm{X}}$ and
$U^{yy}(t)=e^{-it \bm{Y} \bm{Y}}$, which can be transformed from the time evolution operator of an Ising spin pair $U^{zz}(t)=e^{-it \bm{Z} \bm{Z}}$.
The circuit for the operator $U^{zz}(t)$ is composed of a rotation gate $\bm{R_z}(2t)$ and two $\bm{C_X}$ gates,
as shown in Fig.~\ref{fig:exp} (a).  
We can convert the operator $U^{zz}(t)$ into $U^{xx}(t)$ using Hadamard gates via $\bm{H}\bm{Z}\bm{H}=\bm{X}$.
Similarly, we implement  $U^{yy}(t)$ by wrapping the time-evolution operator $U^{zz}(t)$ with Hadamard gates and $\bm{S}$-gates
on each qubit, where $\bm{S}$ is a phase shift gate that rotates a $\pi/2$ radian about the $z$-axis, inducing $\bm{S^\dag}\bm{X}\bm{S}=-\bm{Y}$. 
The circuits for  $U^{xx}(t)$ and $U^{yy}(t)$ are shown in Fig.~\ref{fig:exp} (b) and Fig.~\ref{fig:exp} (c), respectively. 

%%%%%%%%%%%%%%%%% FIG4 %%%%%%%%%%%%%%%%%%%%%%%
\begin{figure}[tb!]
\centerline{\includegraphics[width=6.6cm]{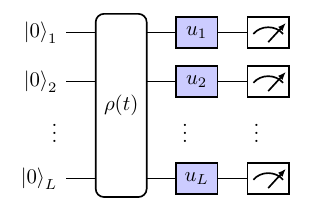}}
%\vskip-2mm
\caption{Randomized measurements implemented on an $L$-qubit state $\rho(t)$ using single-qubit Haar random unitaries $\{\bm{u}_i\}$ followed by a measurement in the computational basis}
\label{fig:random}
%\vskip-3mm
\end{figure}
%%%%%%%%%%%%

\subsection{The R\'enyi entanglement entropy}

The key ingredient of the second-order R\'enyi entropy is the purity
$\text{Tr}[\rho^2]$ of a given (reduced) density matrix.  A recently
developed measurement protocol employing statistical correlations of locally randomized
measurements~\cite{RM,RM_PRA97,RM_PRA,RM_rev} enables us to measure the purity with an adequate accuracy on current noisy quantum devices
via classical postprocessing.
This measurement procedure for our simulation of the  R\'enyi entanglement entropy is as follows: 
(i) Generate a post-quench state, $\rho(t)$, of $L$ qubits at time $t$;
(ii) Draw $N_U$ sets random unitaries of the form $U=u_1\otimes\cdots\otimes u_L$,
where the $u_i$ are single-qubit Haar random unitaries sampled independently from the circular ensemble~\cite{RM_PRA,Haar};
(iii) Apply each set of $U$ to the state $\rho(t)$ and
measure the 
the probabilities $P(\mathbf{s}_{I})$ of obtaining computational basis states $\ket{\mathbf{s}_{I}}=\ket{s_1,\cdots,s_{N_I}}$ 
for the considered subsystem $I$ of $N_I$ qubits.
The purity $\text{Tr}[\rho_I^2]$ for the reduced density matrix is then estimated using
\be
   \text{Tr}[\rho_I^2]=2^{N_I}\sum_{\mathbf{s}_{I},\mathbf{s}'_{I}} (-2)^{-D[\mathbf{s}_{I},\mathbf{s}'_{I}]} \overline{P(\mathbf{s}_{I})P(\mathbf{s}'_{I})}\,,
   \label{eq:purity}
\ee
where $D[\mathbf{s}_{I},\mathbf{s}'_{I}] \equiv \abs{\{j\in I | s_j\neq s'_j \}}$ is the Hamming distance between $\ket{\mathbf{s}_{I}}$ and $\ket{\mathbf{s}'_{I}}$,
and $\overline{\cdots}$ denotes the ensemble average over the $N_U$ sets of random unitaries.
From $\text{Tr}[\rho_I^2]$, we obtain the R\'enyi EE (see Eq.~(\ref{eq:S})). 
This procedure is implemented for each instant of time that we consider to obtain the time-dependent R\'enyi EE $S(t)$. 
In our simulations $N_U=100$ sets of random unitaries and $N_M=2^{12}$ measurements per circuit were used to collect one data point,
unless otherwise stated. 
We completed the measurements and the postprocessing through a parallel automation program~\cite{Code} 
that we have developed for our simulations to  efficiently calculate the entanglement entropy. 

%%%%%%%%%%%%%%%%% FIG5 %%%%%%%%%%%%%%%%%%%%%%%
\begin{figure}[tb!]
\centerline{\includegraphics[width=8.6cm]{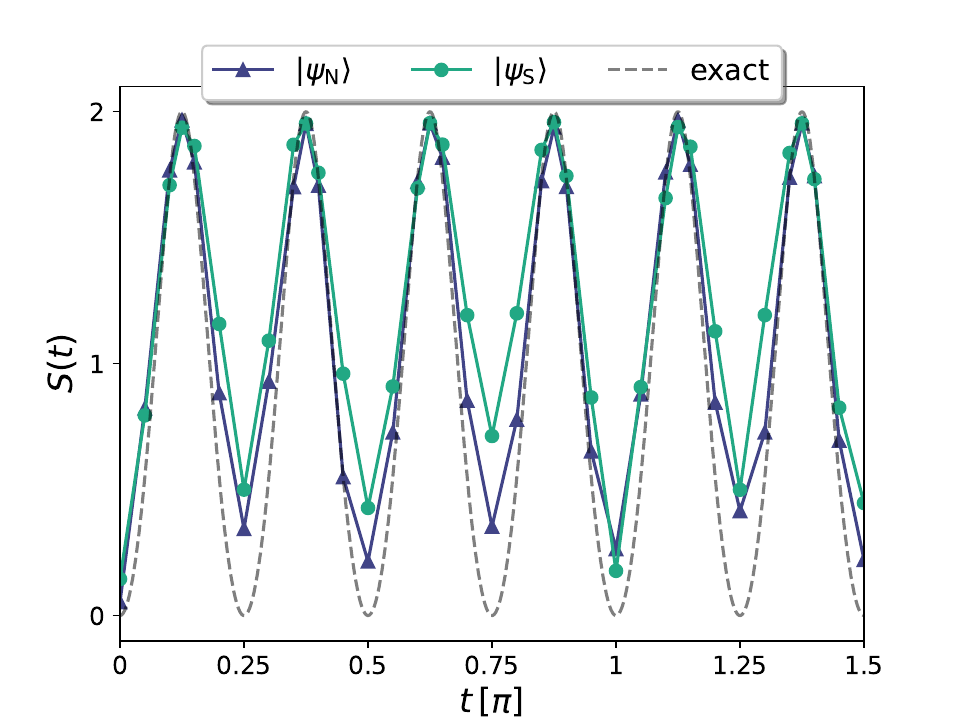}}
%\vskip-2mm
\caption{Results for the half-chain entanglement entropy for a chain with $L=4$ and PBC 
after quenches from the initial states $\ket{\psi_\text{N}}$ and $\ket{\psi_\text{S}}$, simulated on the IBM device {\sl ibm\_mumbai}. The dashed line is the analytical result given in Eq.~(\ref{eq:S01}).}   
\label{fig:L4}
%\vskip-3mm
\end{figure}
%%%%%%%%%%%%

Fig.~\ref{fig:L4} shows the simulation results for short chains with $L=4$ from the IBM devices, compared with exact data.
%For $L=4$, 
The $L=4$ system is a special case in which
the half-chain
entanglement entropies starting from both initial states $\ket{\psi_\text{N}}$ and $\ket{\psi_\text{S}}$
have the same form of time dependence.
Here we observe a clear qualitative agreement between the numerical data from the IBM devices and exact solutions.
In particular, the period of the oscillations and the EE values around the peaks are accurately produced
in the raw data from the IBM device. 
%\comm{Randomized measurements are known to be less accurate for pure states~\ref{RM_PRA97,RM_PRA} 
%because a pure (reduced) density matrix is not a probabilistic mixture of
%other states. As a result, there is an increased deviation from the exact
%solution in the regions where $S\to 0$.}  

%%%%%%%%%%%%%%%%% FIG6 %%%%%%%%%%%%%%%%%%%%%%%
\begin{figure}[tb!]
\centerline{\includegraphics[width=8.6cm]{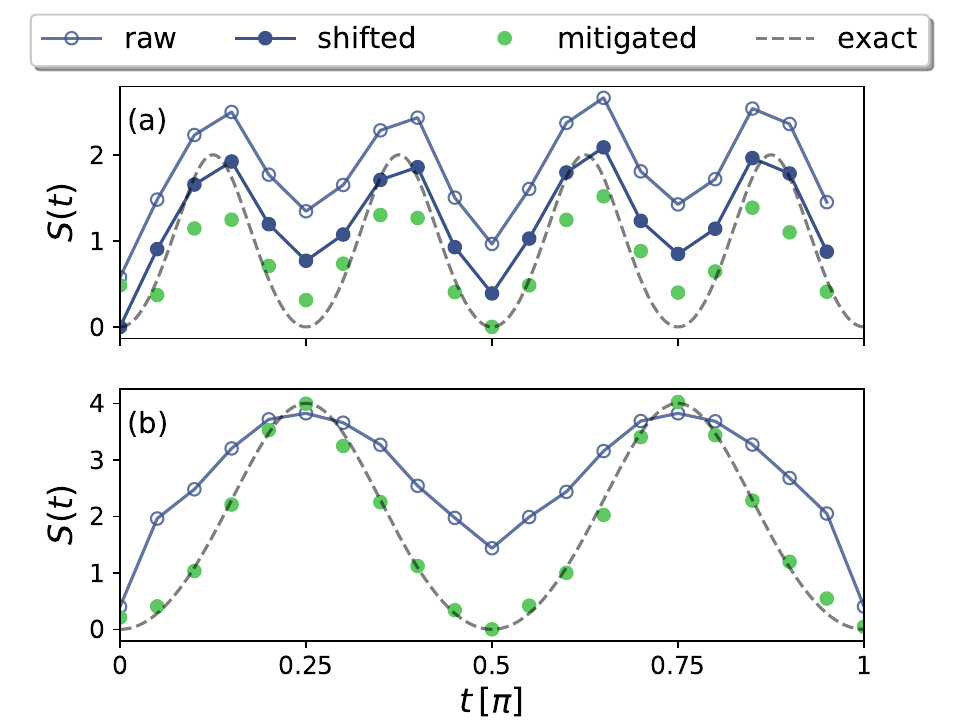}}
%\vskip-2mm
\caption{Results for the half-chain entanglement entropy for a chain with $L=8$ and PBC
after quenches from the initial states $\ket{\psi_\text{N}}$ (a) and $\ket{\psi_\text{S}}$ (b),
simulated on the IBM device {\sl ibm\_montreal}.
The dashed lines are the analytical results given in Eqs.~(\ref{eq:S01}) and (\ref{eq:S02}), respectively.
The simulation results for $\ket{\psi_\text{S}}$ after applying error mitigation using the global depolarizing ansatz
show an excellent match with the exact solution.}
\label{fig:L8_pbc}
%\vskip-3mm
\end{figure}
%%%%%%%%%%%%

%%%%%%%%%%%%%%%%% FIG7 %%%%%%%%%%%%%%%%%%%%%%%
\begin{figure}[tb!]
\centerline{\includegraphics[width=8.6cm]{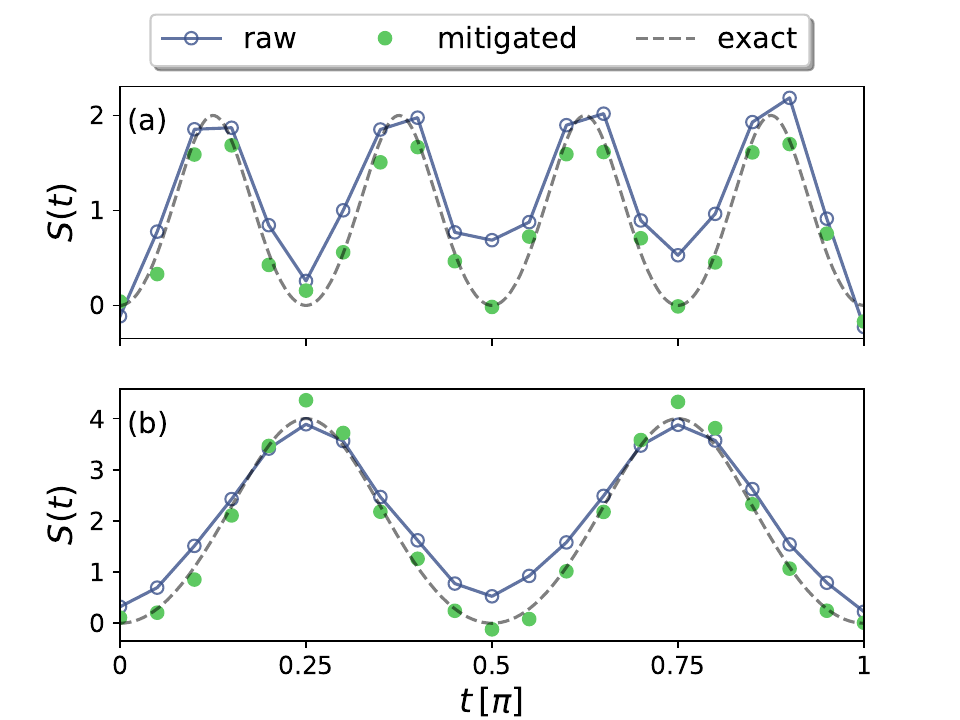}}
%\vskip-2mm
\caption{Results for the entanglement entropy of four bulk qubits in an open chain with $L=8$
after quenches from the initial states $\ket{\psi_\text{N}}$ and $\ket{\psi_\text{S}}$,  simulated on {\sl ibm\_auckland}.
The error-mitigated data are adjusted to align the average value of the valley minima to zero.}
\label{fig:L8_obc}
%\vskip-3mm
\end{figure}
%%%%%%%%%%%%

For longer chains, there is however a serious decline in the accuracy of the simulation results, as shown in Fig.~\ref{fig:L8_pbc} for $L=8$.
For the initial state $\ket{\psi_\text{N}}$, there is a entropy shift when compared to the exact solution.
This offset, possibly due to some inherent error in the device,
can be partially eliminated by aligning the entropy at $t=0$ with the exact value $S(t=0)=0$. 
We note that the shift mitigation approach has also applied in previous studies~\cite{Confinement}.  
To eliminate the deviations beyond the constant shift,
we apply a simple error mitigation strategy using the global depolarizing noise ansatz, along the lines of Ref.~\cite{Mitigation_PRE}.
We assume that the entire quantum state of $L$ qubits under the noise is
\be
    \rho = (1-p_\text{tot})\rho_\text{exact} + \frac{p_\text{tot}}{2^L} \mathds{1}\,,
    \label{eq:noisy}
\ee  
where $p_\text{tot}$ is the total probabilistic error rate and 
$\rho_\text{exact}$ is the exact density matrix without noise.
With the assumption in Eq.~(\ref{eq:noisy}), 
we have the relation between the purities for the noisy density matrix obtained from randomized measurements
and for the exact density matrix:
\be
    \text{Tr}[\rho^2_I] = (1-p_\text{tot})^2\text{Tr}[\rho_{I,\text{exact}}^2]+\frac{p_\text{tot}(1-p_\text{tot})}{2^{N_I-1}} + \frac{p^2_\text{tot}}{2^{N_I}}\,,
\ee 
where $I$ denotes a subsystem of $N_I$ qubits.
The error rate $p_\text{tot}$ can be extracted from $\text{Tr}[\rho^2]$ when the full density matrix for a pure state 
is considered since $\text{Tr}[\rho^2_{\text{exact}}]=1$.
We have determined $p_\text{tot}$ for each time step to carry out the error mitigation for the time-dependent data.
In addition, the mitigated data in Fig.~\ref{fig:L8_pbc} are adjusted to align the minimum to zero to correct negative entropies
in the minima.
The results for $\ket{\psi_\text{S}}$ after applying error mitigation clearly show an excellent match with the exact data,
while the mitigated results for $\ket{\psi_\text{N}}$ with global depolarizing ansatz are merely passable.
We note that the oscillation frequency of the dynamical EE for $\ket{\psi_\text{N}}$ is twice that of 
$\ket{\psi_\text{S}}$. As a result, there are more time points where the exact EE becomes zero 
(i.e., the reduced density matrix corresponds to a pure state) and more intervals where the EE remains close to zero. 
Randomized measurements are known to be less accurate for pure states~\cite{RM_PRA97,RM_PRA}
because a pure (reduced) density matrix is not a probabilistic mixture of
other states. This may explain why the dynamical EE associated with the initial state
$\ket{\psi_\text{N}}$ shows a larger deviation from the exact solution compared to $\ket{\psi_\text{S}}$.
The increased variance for $S\to 0$ is also evident in the raw data presented in Fig.~\ref{fig:L4}.

%%%%%%%%%%%%%%%%% FIG8 %%%%%%%%%%%%%%%%%%%%%%%
\begin{figure}[tb!]
\centerline{\includegraphics[width=6.6cm]{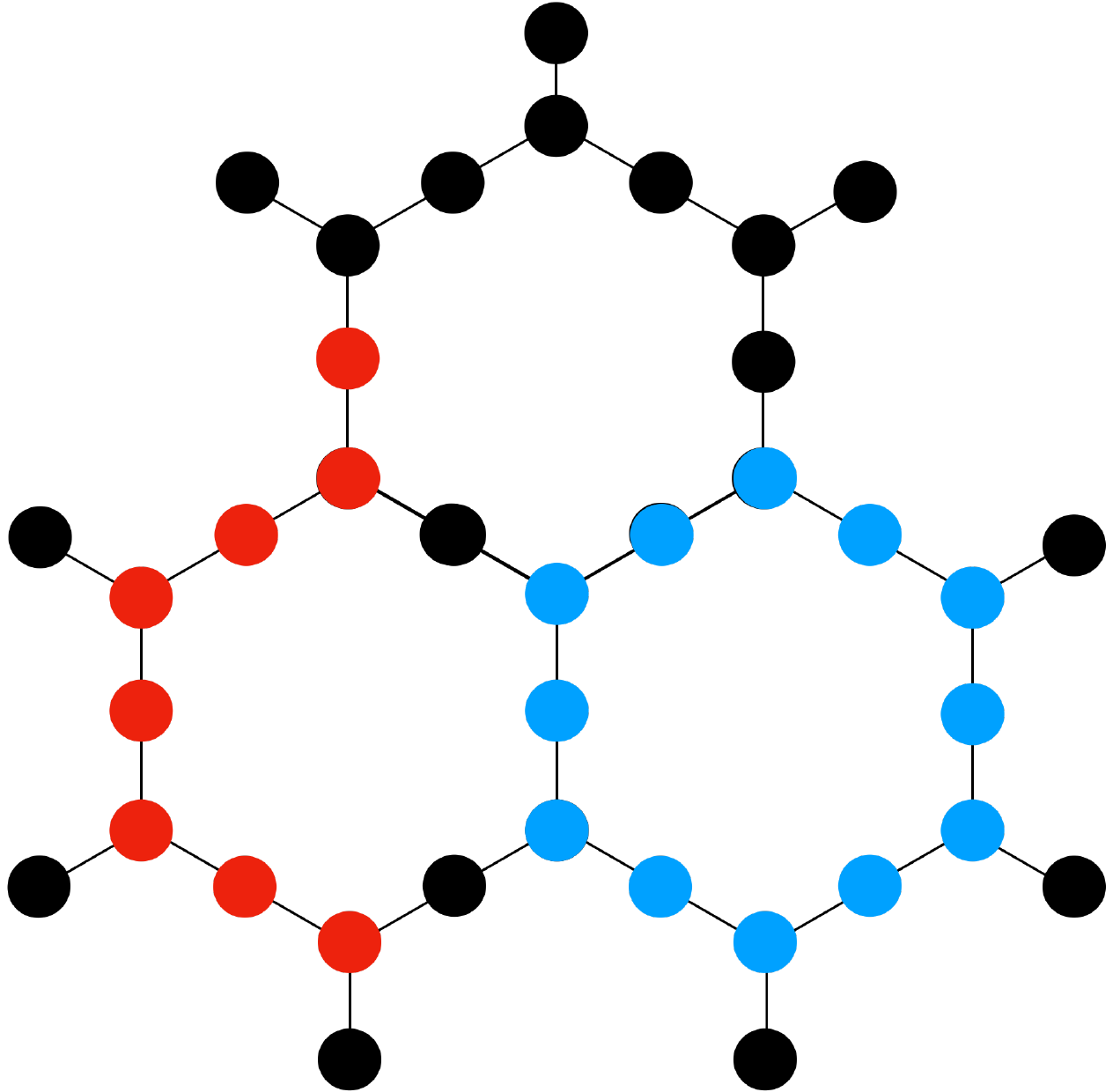}}
%\vskip-2mm
\caption{A schematic plot of the heavy-hexagon lattice, widely adopted in the IBM quantum processors.
Each circle represents a physical qubit. The blue circles form a ring layout and the red circles form a line layout that
are selected in our simulations for a 12-qubit chain and a 8-qubit chain, respectively.}
\label{fig:hex}
%\vskip-3mm
\end{figure}
%%%%%%%%%%%%

%%%%%%%%%%%%%%%%% FIG9 %%%%%%%%%%%%%%%%%%%%%%%
\begin{figure}[tb!]
\centerline{\includegraphics[width=8.6cm]{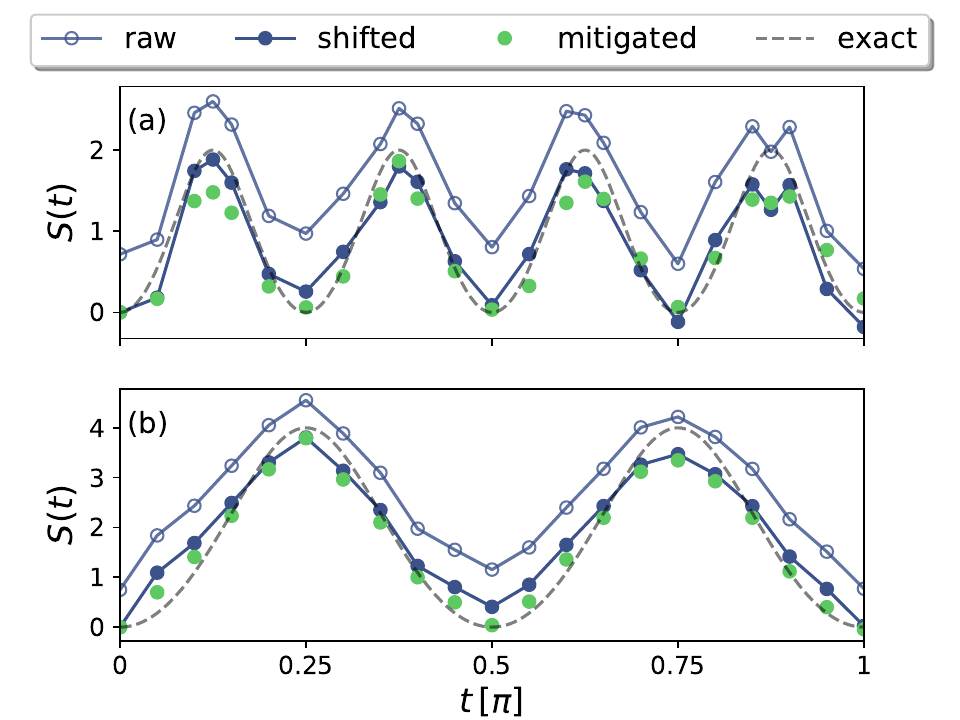}}
%\vskip-2mm
\caption{Results for the half-chain entanglement entropy for a chain with $L=12$ and PBC
after quenches from the initial states $\ket{\psi_\text{N}}$ (a) and $\ket{\psi_\text{S}}$ (b).
The simulations were carried out on {\sl ibmq\_mumbai}.
For $L=12$, we carried out $N_M = 2^{14}$  measurements per circuit to reduce statistical error in the calculation of
the estimator in Eq.~(\ref{eq:purity}).
}
\label{fig:L12_pbc}
%\vskip-3mm
\end{figure}
%%%%%%%%%%%%

The deviation from the exact solutions can have more than one source.
For example, the accuracy of the estimated purity by randomized measurements
depends crucially on the number of random unitaries, $N_U$, and the number of measurements per random unitary, $N_M$; 
as analyzed  in~\cite{RM_PRA97,RM_PRA}, the statistical errors scale with $1/\sqrt{N_U}$ and $1/N_M$, respectively.
%as implied in Eq.~(\ref{eq:purity}).
However, the primary source of error for our case may stem from the PBC imposed on the model, which
requires connectivity between the first and last qubits via ancilla qubits and additional gates,
thereby leading to higher error rates.
For example, the circuit depth for the simulations of the chains with PBC is 40 (45) for the initial state $\ket{\psi_\text{N}}$ ($\ket{\psi_\text{S}}$),
while the depth is reduced to 15 (19) when open chains with OBC are considered.  
We can indeed reproduce the results for $L=8$ with higher accuracy by considering
the EE for a segment of bulk qubits (from the 3rd to the 6th qubit) in an open chain.
The partition has two boundaries, resulting in the EE that is identical to that 
in the PBC case (cf. Appendix~\ref{append}).
The results for the open chain displayed in Fig.~\ref{fig:L8_obc} largely agree quantitatively 
with the exact data, even without applying error mitigation.
After applying the global depolarizing error mitigation, the overall shape of
the EE as a function of time, in particular around the entropy valleys, is corrected. 
However, we need to apply a constant shift to align the entropy at $t=0$ (the minimum entropy) to zero.
%\comm{Another observation is 
%that the oscillation frequency of the dynamical EE for $\ket{\psi_\text{N}}$ is twice that of
%$\ket{\psi_\text{S}}$. As a result, there are more time points where the EE becomes zero (i.e., the reduced density matrix corresponds to a pure state) and more intervals where the EE remains close to zero.
%Randomized measurements are known to be less accurate for pure states~\cite{RM_PRA97,RM_PRA}
%because a pure (reduced) density matrix is not a probabilistic mixture of
%other states. As a result, there is an increased deviation from the exact
%solution in the intervals where $S\to 0$. This may explain why the dynamical EE associated with the initial state
%$\ket{\psi_\text{N}}$ overall shows a larger deviation from the exact solution compared to $\ket{\psi_\text{S}}$.
%The increased variance for $S\to 0$ is also evident in the raw data in Fig.~\ref{fig:L4}, 
%as well as in Fig.~\ref{fig:L8_pbc}.}

In addition to the 4- and 8-qubit systems, we also utilize the IBM heavy-hexagon lattice
to simulate a 12-qubit chain with PBC that forms a ring in the hexagonal arrangement (see Fig.~\ref{fig:hex}).
The circuit depth required in our simulations for the 12-qubit system in this ring layout is the same as for the 8-qubit chain with OBC, namely
15 layers for the initial state $\ket{\psi_\text{N}}$ and 19 layers for $\ket{\psi_\text{S}}$.
Although the number of qubits is increased to 12,
the results for the dynamical entanglement entropy, shown in Fig.~\ref{fig:L12_pbc},
match well with the exact solutions through the simple shift mitigation, and can be further improved with 
the global depolarizing error mitigation.

\subsection{The twist operator}

%%%%%%%%%%%%%%%%% FIG10 %%%%%%%%%%%%%%%%%%%%%%%
\begin{figure}[tb!]
\centerline{\includegraphics[width=8.6cm]{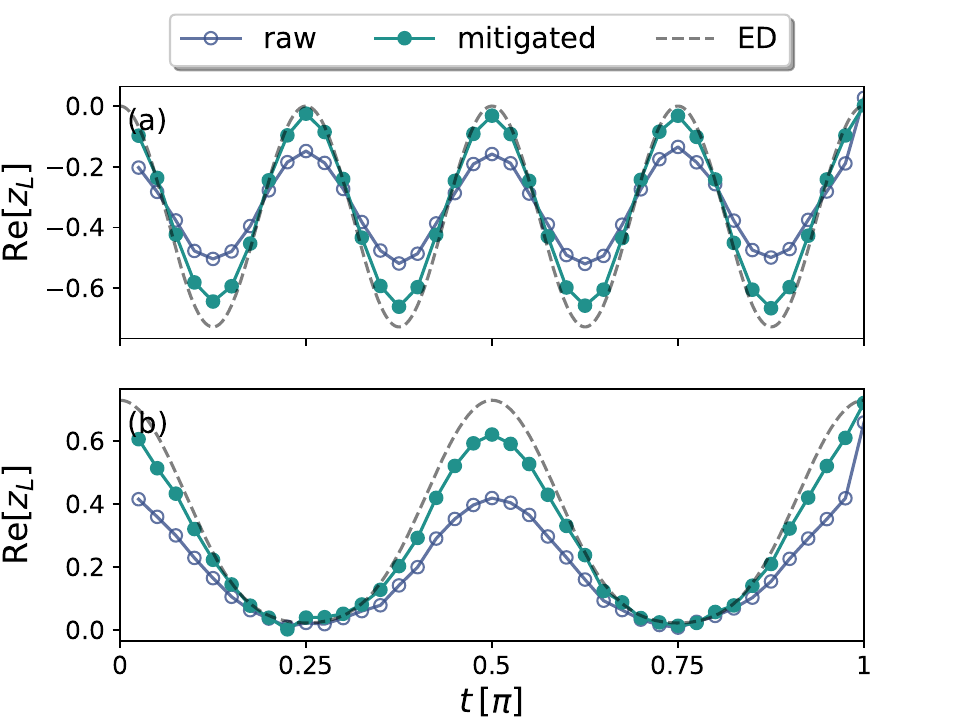}}
%\vskip-2mm
\caption{Results, obtained from {\sl ibm\_auckland}, for the twist order parameter for a chain with $L=8$ and PBC
after quenches from the initial states $\ket{\psi_\text{N}}$ (a) and $\ket{\psi_\text{S}}$ (b).
The dashed lines are results obtained by exact diagonalization (ED).}
\label{fig:L8_zl}
%\vskip-3mm
\end{figure}
%%%%%%%%%%%%

%%%%%%%%%%%%%%%%% FIG11 %%%%%%%%%%%%%%%%%%%%%%%
\begin{figure}[tb!]
\centerline{\includegraphics[width=8.6cm]{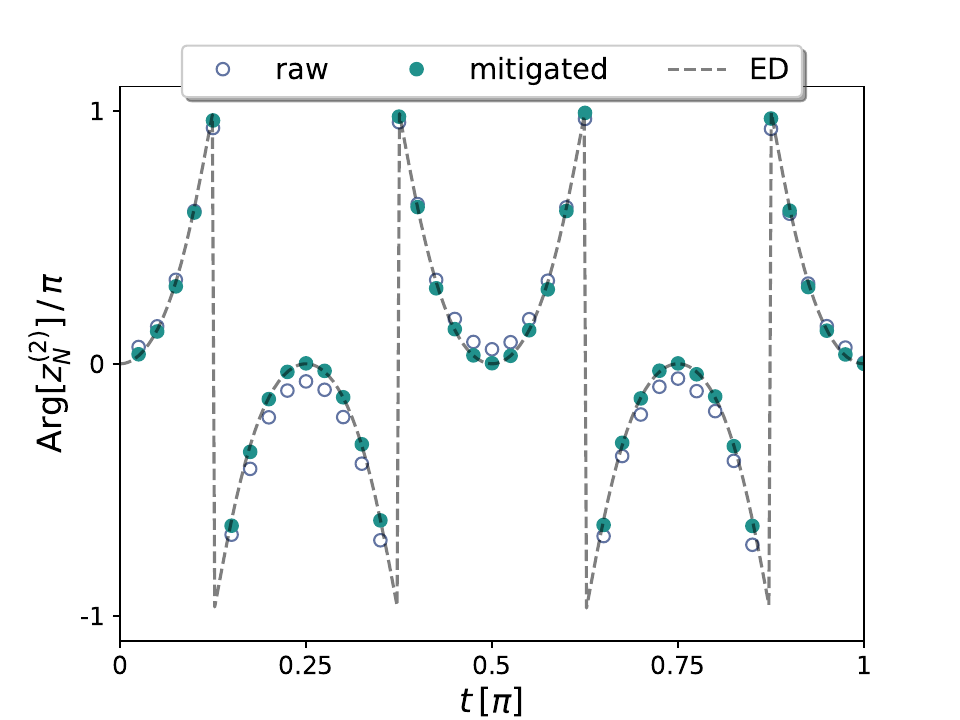}}
%\vskip-2mm
\caption{The dynamical Berry phase for a chain with $L=8$ and PBC after a quench from the initial state $\ket{\psi_\text{N}}$, simulated on {\sl ibm\_auckland}. The dashed lines are results obtained by exact diagonalization.}
\label{fig:L8_B}
%\vskip-3mm
\end{figure}
%%%%%%%%%%%%

The twist operator defined in Eq.~(\ref{eq:twist}) is equivalent to
a product of $\bm{R_z}$ rotation operators: $\otimes_{j=1}^L\bm{R_z}(\theta_j)$ with $\theta_j=-2\pi j/L$
for the angle of rotation on the $j$th-qubit. 
In practice, we can circumvent the accumulated errors resulting from the series of rotation operators by directly 
measuring $\otimes_{j=1}^L\bm{Z}$ for each instant of time, a straightforward process in the computational basis.
Given a set of measured bitstrings $\{\mathbf{s}: \mathbf{s}=s_1s_2\cdots s_L \text{ with } s_j=0$ \text{ or } $1\}$ 
with occurring probabilities $P(\mathbf{s})$,
the twist order parameter $z_L$ (defined in Eq.~(\ref{eq:zL})) is then obtained via postprocessing using
\be
     z_L= \sum_{\mathbf{s}} \exp \left[i\frac{\pi}{L} \sum_{j=1}^L j(1-2s_j) \right] P(\mathbf{s})\,.
     \label{eq:zl_bits}
\ee 

The unmitigated data for $z_L$ in Fig.~\ref{fig:L8_zl}, both for the initial states $\ket{\psi_\text{N}}$ and $\ket{\psi_\text{S}}$, 
display oscillations with periods identical to those of the exact diagonalization (ED) results, but have considerably smaller amplitudes.
We have applied a simple postselection method by only considering bitstrings with an equal number of '0's and '1's,
in accordance with particle number (magnetization) conservation requirements. 
We find an increase in amplitudes for the data after postselecting; in particular, the results for
quenches starting from the N\'eel state $\ket{\psi_\text{N}}$ have shown good quantitative agreement with ED results.

Now we turn to the results for the Berry phase defined in Eq.~(\ref{eq:berry}) with $q=2$.
Similar to $z_L$, the amplitude $z_N^{(q)}$ is obtained by
\be
   z_N^{(2)}= \sum_{\mathbf{s}} \exp \left[i\frac{4\pi}{L} \sum_{j=1}^L j(1-s_j) \right] P(\mathbf{s})\,.
     \label{eq:zn_bits}
\ee
for a  set of bitstrings $\{\mathbf{s}\}$ with occurring probabilities $P(\mathbf{s})$.
The Berry phase is then the argument of this complex amplitude: $\gamma_L=\text{Arg}[z_N^{(2)}]$.
Here we only consider the case for the initial state $\ket{\psi_\text{N}}$ since the case for $\ket{\psi_\text{S}}$
has zero Berry phase in the quench dynamics. 
We have used the same bitstrings from the IBM device for $z_L$ and the same postselected bitstrings
to calculate the amplitude $z^{(2)}_N$. 
In comparison to the twist order parameter $\text{Re}[z_L]$, the Berry phase, as the argument of $z^{(2)}_N$,
appears to be more robust against noise. 
As shown in Fig.~\ref{fig:L8_B}, the Berry phase obtained from the IBM Q device after error mitigation 
is in perfect agreement with exact diagonalization results.

\section{Summary and discussion}
\label{sec:summary}
We have simulated quantum quenches in the SSH chain on NISQ devices provided by
IBM.  We considered particularly simple quench protocols in which the time
evolution is governed by the topological SSH Hamiltonian in the fully dimerized
limit.  This time evolution can be implemented without 
any Trotter error.  Our simulations on the IBM-Q devices demonstrated periodic
oscillations in entanglement dynamics after a quench from a topologically
trivial state, reflecting periodic transitions between two phases with
distinct topological structures.  We have also detected such dynamical
topology using the twist order parameter and the Berry phase, which can be
measured on a quantum device without the need for additional gates.  
Through some error mitigation techniques, we were able to deduce time-dependent
data that match exact solutions for up to 12 qubits. 

The layout of qubits significantly impacts the accuracy of our simulation
results. Specifically, simulating a 8-qubit chain with PBC requires more than
twice as many circuit layers compared to a 8-qubit chain with OBC, leading to
increased noise. On the other hand, the IBM heavy-hex lattice, which consists of
12-qubit rings with nearest-neighbor couplings, provides an ideal platform for
our simulations of 12-qubit chains with PBC using shallow circuits.

To mitigate noise on NISQ computers, we applied postselection based on symmetry
conservation to twist order parameters and the Berry phase. The mitigated
results agree very well with the exact solutions.  However, this simple
mitigation method cannot be applied to the entanglement entropy because the
randomized measurement protocol requires the application of random unitary
gates that break symmetry constraint.  In some cases, the overall structure of
the unmitigated time-dependent entanglement entropy is correct except for a
constant deviation from the exact solution, which can be corrected by applying
a constant shift to align the entropy at $t=0$ to zero (the theoretical value).
In some cases, the unmitigated entanglement entropy as a function of time is
misshapen and the largest amplitude of the oscillation is smaller than the
theoretical value; we were able to correct the shape and the amplitudes through
a global depolarizing error mitigation method~\cite{Mitigation_PRE}. However, a
constant shift to align the minimum entropy to zero is often required after
applying the global depolarizing error mitigation.  By combining global
depolarizing error mitigation and shift mitigation, we were able to achieve
mitigated results that are in excellent agreement with the exact solutions.

The LSM twist operator applied in our simulations can be used to formulate
various quantities and concepts that are of fundamental importance in 
condensed matter physics, 
such as order parameters to characterize gapped
phases~\cite{Twist,Hatsugai_PRB77}, Berry phases~\cite{Position}, polarization~\cite{Position}
and localization length~\cite{Localization}.
Recently, the applications of the LSM twist
operator have also been generalized to open quantum
systems~\cite{LSM_Ryu,LSM_open}.  The twist operator is diagonal in the
computational basis on  quantum computers.  This property  allows us to
calculate its related quantities on classical computers by processing the
bitstrings and their corresponding probabilities from quantum device
measurements, without incurring additional overhead in quantum hardware. The
method makes many interesting applications of the twist operator  amenable to
simulations on NISQ devices.

A natural extension of this work is to simulate the quench dynamics governed by the topological
Hamiltonian with partial dimerization, i.e. with $(J'>J\neq 0)$.  In the partially
dimerized case, the time evolution of the entanglement entropy 
has a more complex pattern. It exhibits slow oscillations with large amplitudes, which modulate faster and smaller oscillations~\cite{Eisler}.
One of the main challenges in simulating such quench dynamics using
conventional Trotter decomposition
on digital quantum computers is the need for deep quantum circuits to implement Trotterization, 
which often significantly reduces the accuracy of quantum simulations on current NISQ hardware.  
Nevertheless, various strategies have been proposed to
reduce circuit depths for time evolution~\cite{Compression_PRA2022,Chiral_dyn,Time_Pollmann,Lee_npjQ,Heyl,Trotter_mn}. 
It will be interesting to implement and extend these approaches on quantum computers to realize the complex dynamics of
quantum entanglement and topology in the partially dimerized SSH model.

\begin{acknowledgments}
        The authors acknowledge support from the IBM Q Hub at National Taiwan University, the National Science and Technology Council (NSTC) and the National Center for Theoretical Sciences (NCTS) in Taiwan. This work was supported under Grants No.~NSTC 111-2119-M-007-009, 112-2119-M-007-008 and 112-2112-M-004-006.
\end{acknowledgments}

\appendix
\onecolumngrid
\section{Derivation of the entanglement entropy}
\label{append}
In this appendix, we present the analytical derivation of the entanglement entropies of the post-quench states.
%For the free-fermion model considered in this paper,
\subsection{Correlation matrix method for free-particle models}
We first recall the correlation matrix method for calculating the entanglement entropy of
a noninteracting system proposed by Peschel~\cite{Peschel}. 
For a free-fermion system described by a quadratic Hamiltonian in terms of fermionic operators $c_j$ and $c_j^\dag$,
the reduced density matrix $\rho_I$ for a subsystem $I$ can be written in the form 
of a thermal density matrix with an entanglement Hamiltonian $\mathcal{H}_I$:
\be
	\rho_I=\frac{1}{Z} e^{-\mathcal{H}_I}, \quad \text{with } Z=\text{Tr}[e^{-\mathcal{H}_I}]\,,
        %,\quad \mathcal{H}_I=\sum_{i\in I} \varepsilon_i f_i^\dag f_i\,,
\ee 
where the entanglement Hamiltonian is quadratic in the fermionic operators and can be diagonalized in the form:
\be
	\mathcal{H}_I=\sum_{\ell\in I} \varepsilon_\ell f_\ell^\dag f_\ell\,,
\ee
with $c_j=\sum_\ell U_{j\ell}f_\ell$. The single-particle correlation function $C_{ij}=\e{c_i^\dag c_j}$ 
in subsystem $I$
can be constructed from the reduced density matrix $\rho_I$:
\be
	C_{ij} = \text{Tr}[\rho_I c_i^\dag c_j] = \sum_\ell U_{i\ell}^* U_{j\ell}\frac{1}{1+e^{\varepsilon_\ell}}\,.
\ee
Thus, the correlation matrix $C$ 
has eigenvalues $\xi_\ell$ that are related to the eigenvalues $\varepsilon_\ell$ of the entanglement Hamiltonian via
\be
       \xi_\ell = \frac{1}{1+e^{\varepsilon_\ell}}\,.      
\ee
The $\alpha$th-order R\'enyi entanglement entropy of the subsystem $I$, defined as
\be
         S^{(\alpha)}=\frac{1}{1-\alpha} \log_2 \text{Tr}[\rho_I^\alpha]\,,
\ee
is explicitly given by
\be
        S^{(\alpha)} = \frac{1}{1-\alpha} \sum_{\ell \in I} \left[\log_2(1+e^{-\alpha \varepsilon_\ell}) -
        \log_2(1+e^{-\varepsilon_\ell})^\alpha\right]\,,
\ee
or can be written in terms of the eigenvalues $\xi_\ell$ of the correlation matrix as
\be
         S^{(\alpha)}= \frac{1}{1-\alpha} \sum_{\ell \in I} \log_2 [(1-\xi_\ell)^\alpha+\xi_\ell^\alpha]\,.
\ee
For the case with $\alpha=2$, we obtain Eq.~(\ref{eq:S_cor}).

\subsection{Time-dependent Wannier states}
Consider an initial state in momentum space, 
$\psi_k(0) = \left(\alpha a_k^{\dagger}+\beta b_k^{\dagger}\right)|\emptyset\rangle$, where
$|\emptyset\rangle$ is the vacuum state,
$a_k^\dagger(b_k^{\dagger})$ denotes the creation operator at $A(B)$-sublattice site with momentum $k$,
and  $\alpha, \beta$ are coefficients satisfying $|\alpha|^2+|\beta|^2=1$.
The post-quench state governed by the SSH Hamiltonian defined in Eqs.~(\ref{eq:Hk}) and (\ref{eq:d}) is
\begin{eqnarray}\label{eq:ut}
        |\psi_k(t)\rangle&=&\Big\{\left[\alpha\cos(d_kt)-i\beta\sin(d_kt)\hat{d}_-\right] a_k^{\dagger}\nonumber\\
        &+&\left[\beta\cos(d_kt)-i\alpha\sin(d_kt)\hat{d_+}\right]
        b_k^{\dagger}\Big\}|\emptyset\rangle,
\end{eqnarray}
where $d_k=|\mathbf{d}_k|$ and $\hat{d}_{\pm}$ is defined as $\hat{d}_{\pm}=(d_k^x\pm id_k^y)/d_k=e^{\pm i\phi}$ with $\phi=\tan^{-1}(d_k^y/d_k^x)$. In the flat band limit with $(J=0, J'=1)$, we have $\hat{d}_{\pm}=e^{\pm i k}$;
for the case with $(J=1, J'=0)$, it is $\hat{d}_{\pm}=1$. Here we focus on the case $(J=0, J'=1)$ as considered
in this paper.   

Applying the inverse Fourier transform to the state in Eq.~(\ref{eq:ut}), 
we obtain the Wannier state at the $m$th cell~\cite{Asboth2016,Vanderbilt2018book},
\begin{eqnarray}\label{eq:wannier}
        |w(m,t)\rangle&=&\frac{1}{\sqrt{N}}\sum_ke^{ikm}|\psi_k(t)\rangle\nonumber\\
        &=&\Big\{\cos(d_kt)\left[\alpha a_m^{\dagger}+\beta b_m^{\dagger}\right] \nonumber\\
        &-&i\sin(d_kt)\left[\beta a_{m_+}^{\dagger}+\alpha b_{m_-}^{\dagger}\right]
        \Big\}|\emptyset \rangle,
\end{eqnarray}
where we have introduced $m_{\pm}=m\pm 1$ to denote the neighboring cells, and applied the relation 
$\frac{1}{N}\sum_ke^{ik(m-\ell)}=\delta_{\ell,m}$. 
%The post-quench state given in Eq.~(\ref{eq:wannier})
%spans three lattice spacings from the $(m-1)$th to $(m+1)$th cells. 

\subsection{Quench from the N\'eel state $\ket{\psi_\text{N}}$}
For the initial state $\ket{\psi_\text{N}}$ defined in Eq.~(\ref{eq:psi_N}), we have $\alpha=1, \beta=0$.
The Wannier state after the quench is then
\begin{eqnarray}\label{eq:wN}
        |w_\text{N}(m,t)\rangle=\left[\cos(d_kt) a_m^{\dagger}
        -i\sin(d_kt) b_{m_-}^{\dagger}\right]
        |\emptyset\rangle\,.
\end{eqnarray}
%We consider a symmetric bipartition with a boundary between the $(m-1)$-th and $m$-th cells. 
The correlation matrix associated with $\ket{w_\text{N}(m,t)}$ is given by
%\begin{widetext}
\begin{eqnarray}
        \begin{pmatrix}
                \langle b_{m_-}^{\dagger}b_{m_-}\rangle &       \langle b_{m_-}^{\dagger}a_{m}\rangle\\
                        \langle a_{m}^{\dagger}b_{m_-}\rangle&\langle a_{m}^{\dagger}a_{m}\rangle
        \end{pmatrix}=
        \begin{pmatrix}
        \sin^2(d_kt) & -i\sin(d_kt)\cos(d_kt)\\
        i\sin(d_kt)\cos(d_kt)& \cos^2(d_kt)
        \end{pmatrix}.
\end{eqnarray}
%\end{widetext}
The eigenvalues of this correlation matrix are $0$ and $1$, resulting in zero EE. This  indicates that
the Wannier state does not contribute to the EE if it lies entirely within the subsystem.
Here we consider a symmetric bipartition with a boundary between the $(m-1)$-th and $m$-th cells. 
The submatrix of the left half is $\e{b_{m_-}^{\dagger}b_{m_-}}$, with an eigenvalue $\sin^2(d_kt)$.
The second-order R\'enyi EE of the left half is then
\begin{eqnarray}
        S_\text{OBC}&=&-\log_2\left[
        \sin^4(d_kt)+\cos^4(d_kt)
        \right]\nonumber\\
        &=&-\log_2\left[
        1-\sin^2(2d_kt)/2
        \right]\,,
        \label{eq:sneel}
\end{eqnarray}
where the subscript 'OBC' denotes open boundary conditions.
In a chain with PBC, there are two boundaries between the partitions, with each boundary intersecting a
Wannier state. Consequently, the R\'enyi EE for PBC is twice that of OBC.
\be
   S_\text{PBC} = -2 \log_2\left[1-\sin^2(2d_kt)/2 \right]\,.
\ee  

\subsection{Quench from the fully dimerized state $\ket{\psi_\text{S}}$}

For a quench from the fully dimerized state $\ket{\psi_\text{S}}$, the post-quench Wannier state
with $\alpha=1/\sqrt{2}, \beta=-1/\sqrt{2}$ becomes
\begin{eqnarray}\label{eq:wS}
        |w_\text{S}(m,t)\rangle&=&\frac{1}{\sqrt{2}}\Big\{\cos(d_kt)\left[ a_m^{\dagger}- b_m^{\dagger}\right] \nonumber\\
        &+&i\sin(d_kt)\left[ a_{m_+}^{\dagger}- b_{m_-}^{\dagger}\right]
        \Big\}|\emptyset\rangle\,,
\end{eqnarray}
where the second term  involves two next-nearest-neighbor unit cells.

We consider a bipartition with a boundary located between the unit cells $m$ and $m+1$.
Two Wannier states are associated with this boundary: one formed between cells $m-1$ and $m$, and the other
between cells $m$ and $m+2$.
Thus, we need to include these two Wannier states $|w_\text{S}(m,t)\rangle$ and $|w_\text{S}(m+1,t)\rangle$
to construct the relevant portion of the correlation matrix:
%\begin{widetext}
\begin{eqnarray}
        \begin{pmatrix}
                %a_{m_-}^{\dagger}a_{m_-}\rangle &      \langle a_{m_-}^{\dagger}b_{m_-}\rangle&\langle a_{m_-}^{\dagger}a_{m}\rangle&\langle a_{m_-}^{\dagger}b_{m}\rangle\\
                        \langle b_{m_-}^{\dagger}b_{m_-}\rangle&\langle b_{m_-}^{\dagger}a_{m}\rangle&\langle b_{m_-}^{\dagger}b_{m}\rangle\\
                \langle a_{m}^{\dagger}b_{m_-}\rangle&\langle a_{m}^{\dagger}a_{m}\rangle&\langle a_{m}^{\dagger}b_{m}\rangle\\
                \langle b_{m}^{\dagger}b_{m_-}\rangle&
                \langle b_{m}^{\dagger}a_{m}\rangle&\langle b_{m}^{\dagger}b_{m}\rangle
        \end{pmatrix}=
\frac{1}{2}
\left(
\begin{array}{cccc}
%       0 & 0 & 0 & 0 \\
         {\sin ^2(d_k/t)} & - i \sin (d_kt) \cos (d_kt) &  i \sin (d_kt) \cos (d_kt) \\
          i \sin (d_kt) \cos (d_kt) & {\cos ^2(d_kt)} & - \cos ^2(d_kt) \\
         - i \sin (d_kt) \cos (d_kt) & - \cos ^2(d_kt) & 1 \\
\end{array}
\right).
\end{eqnarray}
%\end{widetext}
The eigenvalues of this matrix are $0, (1-\cos(d_kt))/2$ and $(1+\cos(d_kt))/2$. Thus,
the second-order R\'enyi EE with a single partition boundary, as in an open chain, is given by
\begin{eqnarray}
        S_\text{OBC}=-2\log_2\left[1-\sin^2(d_kt)/2\right].
        \label{eq:ssinglet_obc}
\end{eqnarray}
In a chain with PBC, two boundaries separate the partitions, doubling the R\'enyi EE:
\begin{eqnarray}
        S_\text{PBC}=-4\log_2\left[1-\sin^2(d_kt)/2\right].
        \label{eq:ssinglet_pbc}
\end{eqnarray}

%\subsection{Quench in a short chain of two unit cells}
%In this subsection, we show that the EE after a quench from a fully dimerized state in a short chain 
%of two unit cells is the same of that from a N\'eel state.
%For a chain with only two cells $m=0$ and $m=1$, the Wannier state in Eq.~(\ref{eq:wS}) is reduced to 
%\begin{eqnarray}
%        |w_\text{{S}}(m=0,t)\rangle_{N=2}&=&\frac{1}{\sqrt{2}}\Big\{\cos(d_kt)\left[ a_0^{\dagger}- b_0^{\dagger}\right]
%\nonumber\\
%        &+&i\sin(d_kt)\left[ a_{1}^{\dagger}- b_{1}^{\dagger}\right]
%        \Big\}|\emptyset\rangle,
%\end{eqnarray} 
%since $m_-$ and $m_+$ refer to the same unit cell due to the periodic boundary condition.
%The correlation matrix for the left partition containing the cell $m=0$ is
%%\begin{widetext}
%\begin{eqnarray}
%       \begin{pmatrix}
%		\langle a_0^{\dagger}a_0\rangle &\langle a_0^{\dagger}b_0\rangle\\
%	        \langle b_0^{\dagger}a_0\rangle &\langle b_0^{\dagger}b_0\rangle
%        \end{pmatrix}= 
%        \frac{1}{2}
%        \begin{pmatrix}
%                {\cos ^2(d_kt)} & - \cos ^2(d_kt) \\
%                - \cos ^2(d_kt) & {\cos ^2(d_kt)} 
%	\end{pmatrix}
%\end{eqnarray}
%\end{widetext}
%with eigenvalues $\cos^2(d_kt)$ and $0$. Thus, the R\'enyi EE is 
% \begin{eqnarray}
%        S_{N=2}&=&-\log_2\left[
%        \cos^4(d_kt)+\sin^4(d_kt)
%        \right]\nonumber\\
%        &=&-\log_2\left[
%        1-\sin^2(2d_kt)/2
%        \right],
% \end{eqnarray}
%identical to the EE for the N\'eel state (Eq.~\ref{eq:sneel}).

\twocolumngrid

\end{document}